\documentclass[11pt,aps,fleqn,superscriptaddress,notitlepage,nofootinbib,preprintnumbers,nobalancelastpage]{revtex4-1}
\pdfoutput=1
\usepackage{amsmath,amssymb,graphicx,xspace,tikz}

\usepackage{hyperref}
\usepackage{bm}
\usepackage{epsfig,amsmath,amssymb,verbatim,mathrsfs,hyperref}
\usepackage{xspace}
\usepackage{xcolor}
\usepackage{slashed}
\usepackage{dcolumn}
\usepackage{multirow}
\usepackage{graphicx}
\usepackage{caption}
\usepackage{subcaption}
\usepackage{ulem}
\graphicspath{{./}{Figs/}}

\begin{document}

\newcommand\snowmass{\begin{center}\rule[-0.2in]{\hsize}{0.01in}\\\rule{\hsize}{0.01in}\\
\vskip 0.1in Submitted to the  Proceedings of the US Community Study\\ 
on the Future of Particle Physics (Snowmass 2021)\\ 
\rule{\hsize}{0.01in}\\\rule[+0.2in]{\hsize}{0.01in} \end{center}}

%%%%%%%%%%%%%%%%%%%%%%%%%%%%%%%%%%%%%%%%%%%%%%%%%%

\def\beq{\begin{eqnarray}}
\def\eeq{\end{eqnarray}}
\def\bea{\begin{eqnarray}}
\def\eea{\end{eqnarray}}
%%%%%%%%%%%%%%%%%%%%%%%%%%%%%%%%%%%%%%%%%%%%%%%%%

\title{%Snomwass Whitepaper\\
Probing the Electroweak Phase Transition with Exotic Higgs Decays}
\author{Marcela Carena}
\affiliation{Theoretical Physics Department, Fermi National Accelerator Laboratory, Batavia, Illinois, 60510, USA}
\affiliation{Enrico Fermi Institute, University of Chicago, Chicago, Illinois, 60637, USA}
\affiliation{Kavli Institute for Cosmological Physics, University of Chicago, Chicago, Illinois, 60637,
USA}
\author{Jonathan Kozaczuk}
\affiliation{Department of Physics, University of California, San Diego, La Jolla, CA 92093, USA}
\author{Zhen Liu}
\affiliation{School of Physics and Astronomy, University of Minnesota, Minneapolis, MN 55455, USA}
\author{Tong Ou}
\affiliation{Enrico Fermi Institute, University of Chicago, Chicago, Illinois, 60637, USA}
\author{Michael~J.~Ramsey-Musolf}
\affiliation{Tsung-Dao Lee Institute and  School of Physics and Astronomy, Shanghai Jiao Tong University,
800 Dongchuan Road, Shanghai, 200240 China}
\affiliation{Amherst Center for Fundamental Interactions, Department of Physics, University of Massachusetts, Amherst, MA 01003, USA}
\affiliation{Kellogg Radiation Laboratory, California Institute of Technology,
Pasadena, CA 91125 USA}
\author{Jessie Shelton}
\affiliation{Illinois Center for the Advanced Study of the Universe, Department of Physics, University of Illinois at Urbana-Champaign, Urbana, IL 61801, USA}
\author{Yikun Wang}
\affiliation{Walter Burke Institute for Theoretical Physics, California Institute of Technology, Pasadena, CA 91125, USA}
% \affiliation{Theoretical Physics Department, Fermi National Accelerator Laboratory, Batavia, Illinois, 60510, USA}
% \affiliation{Enrico Fermi Institute, University of Chicago, Chicago, Illinois, 60637, USA}
\author{Ke-Pan Xie}
\affiliation{Department of Physics and Astronomy, University of Nebraska, Lincoln, NE 68588, USA}

%\clearpage
\begin{abstract}
An essential goal of the Higgs physics program at the LHC and beyond
%proposed future colliders 
is to explore the nature of the Higgs potential and shed light on the mechanism of electroweak symmetry breaking. An important class of  models defining the strength and order of the electroweak phase transition is driven by the Higgs boson coupling to a light new state. This Snowmass white paper points out the existence of a region of parameter space where a strongly first order  electroweak phase transition is compatible with exotic decays of the SM-like Higgs boson. A dedicated search for exotic Higgs decays can actively explore this framework  at the Large Hadron Collider (LHC), while 
future exotic Higgs decay searches at the high-luminosity LHC and future Higgs factories will be vital to conclusively probe the scenario.
%Cup substantialImprovement in rich can be provided byFuture A large room for improvement can be potentially achieved via further phenomenological and experimental studies.  
\snowmass
\end{abstract}
\maketitle

\section{Motivation}
%{\flushleft  {\color{blue} [Marcela]}}

An important goal of the Higgs physics program at the LHC and proposed future colliders 
is to explore the nature of the Higgs potential and shed light on the mechanism of electroweak symmetry breaking (EWSB). Given our current understanding of the Higgs boson, we do not yet know how EWSB occurred in the early Universe. The Standard Model (SM) predicts EWSB as a smooth thermodynamic crossover. However, it may well be that new physics beyond the SM (BSM) alters this picture, enabling the possibility of 
%resulting in 
a first-order electroweak phase transition (EWPT). A first-order EWPT could have supplied one of the necessary ingredients for generating the observed baryon asymmetry of the Universe through the mechanism of electroweak baryogenesis~\cite{Morrissey:2012db}. In addition, such a process produces a
% an observable 
stochastic gravitational-wave (GW) 
% background
signature that could be observed in future GW probes~\cite{Caprini:2015zlo, Caprini:2019egz}. A
first-order EWPT could have affected the abundance of primordial relics such as dark matter with masses at or above the electroweak scale~\cite{Wainwright:2009mq}. Beyond these important cosmological implications, mapping out the phase diagram of the electroweak sector is an important undertaking in its own right, analogous to determining the phase diagram of QCD. Understanding the history  of the EWPT can be game changing in our understanding of nature,
%For these reasons and many more, the high-energy physics community needs to explore the nature of the EWPT as thoroughly and robustly as possible in
and present  and future collider experiments have  a crucial role to play in such exploration.

Many BSM scenarios predicting a first-order electroweak phase transition are being tested at the LHC. An important class of such models is that in which the EWPT is driven first-order by the Higgs coupling to a new {\em light} particle~\cite{Profumo:2007wc}. As argued in Ref.~\cite{Kozaczuk:2019pet}, if the new degree of freedom is below $\sim m_Z$, it must be a singlet-like scalar denoted as $s$.  The mixing angle ($\sin\theta$) of this singlet with the SM Higgs boson  ($h$) control the direct production of such scalars at colliders, and can be small or absent entirely if additional symmetries are present. Therefore, these light new particles can be difficult to detect directly. Refs.~\cite{Kozaczuk:2019pet, Carena:2019une} recently emphasized that such scalars can still have a dramatic impact on the EWPT, provided there is a sizable scalar coupling to the SM-like Higgs field.
% is large enough. 
This coupling also controls the $h\rightarrow ss$ branching ratio when $m_s<m_h/2$.  Therefore, these scalars provide a compelling target for exotic Higgs decay searches, both at the LHC and future colliders.

Exotic Higgs decays are a cornerstone of the discovery program at both current and future colliders \cite{Curtin:2013fra,Abada:2019lih,CEPCStudyGroup:2018ghi,Benedikt:2018csr,Abada:2019zxq,Cepeda:2021rql}.  At the HL-LHC, detector upgrades, new trigger and analysis strategies, and increased datasets will steadily increase the sensitivity to small branching ratios, especially in subdominant but cleaner final states  (notably $h\to ss\to bb \tau \tau$, $h\to$ invisible) \cite{deFlorian:2016spz,Cepeda:2019klc}.  Meanwhile, proposed $e^+ e^-$ colliders offer lower integrated luminosities but substantially lower backgrounds, resulting in excellent sensitivity to challenging all-hadronic modes, notably $h\to ss\to bbbb$ \cite{Liu:2016zki}. This white paper aims to establish how searches for exotic Higgs decays can inform our understanding of the early Universe. It will establish parameter space portions compatible with exotic Higgs decays and first-order phase transitions and provide clear and achievable targets for current and future colliders.

\section{Higgs exotic decay target from EWPT considerations}
\label{sec:theory}
%{\flushleft  {\color{blue} [Yikun, Jessie]}}

For strongly first-order phase transitions driven by {\em light} degrees of freedom, the most exciting possibility is a singlet scalar $s$, which can affect the Higgs potential at tree-level.
The existence of experimentally viable parameter space where a light singlet-like scalar can drive the EWPT to be strongly first-order was recently demonstrated in ~Refs.~\cite{Carena:2019une,Kozaczuk:2019pet}. It is useful to classify the possibilities according to whether or not the light singlet scalar respects the $Z_2$ symmetry taking $s\to -s$. Theoretically, this $Z_2$ symmetry, whether or not it is spontaneously broken, can be a useful proxy for a bigger (gauge or global) symmetry, as can be the case when $s$ is part of a hidden sector, as motivated, e.g., by dark matter model-building. Practically, imposing a $Z_2$ symmetry reduces the number of free parameters. Most consequentially, in models with an unbroken $Z_2$, $s$ cannot mix with the SM Higgs. At the same time, if the $Z_2$ is spontaneously broken, the $s$-$h$ mixing angle $\sin\theta$ is related to the exotic branching fraction Br($h\to ss$). In the absence of the $Z_2$ symmetry, $\sin\theta$ and Br($h\to ss$) are independent parameters.  
 Ref.~\cite{Carena:2019une} studies the $Z_2$ spontaneously-broken scenario, while Ref.~\cite{Kozaczuk:2019pet} studies the $Z_2$ unbroken and non-symmetric scenarios.

%the singlet extended SM with spontaneous $Z_2$ breaking. Such a novel scenario of the SM singlet extension is theoretically motivated by the Higgs portal interactions between the SM Higgs doublet and dark Higgs scalars of hidden sectors. Hidden sectors with dark gauge symmetries are well-motivated and extensively studied in dark matter model building. In those models, the dark gauge symmetry is usually spontaneously broken by one or more dark Higgs scalars, providing mass(es) to the dark gauge boson(s). We investigated if such Higgs portal interactions can modify the EW vacuum structure and trigger a first-order EWPhT, studying which, in most cases, the joint scalar sector can be simplified as the SM Higgs ($H$) plus a real singlet scalar ($s$) with a spontaneous breaking $Z_2$ symmetry.
% On the other hand, the dark Higgs can couple to the SM Higgs doublet via scalar portal interactions. It is then interesting to investigate if such interactions can modify the EW vacuum structure and trigger a first-order EWPhT. Although there are various models on dark gauge symmetry and Higgs, when considering the phase transition structure, in most cases the joint scalar sector can be approximated as the SM Higgs ($H$) plus a real singlet scalar ($s$) with a spontaneous breaking $Z_2$ symmetry.

%Although there have been many studies on the EWPT in real singlet extended SM, the case of a spontaneous breaking $Z_2$ distinguishes from the previous researches on strictly $Z_2$-preserving or the general $Z_2$-explicit breaking extension scenarios on the impact on EWPT. 

\begin{figure}
     \centering
     \begin{subfigure}[t]{0.32\textwidth}
        % \centering
         \includegraphics[width=\textwidth]{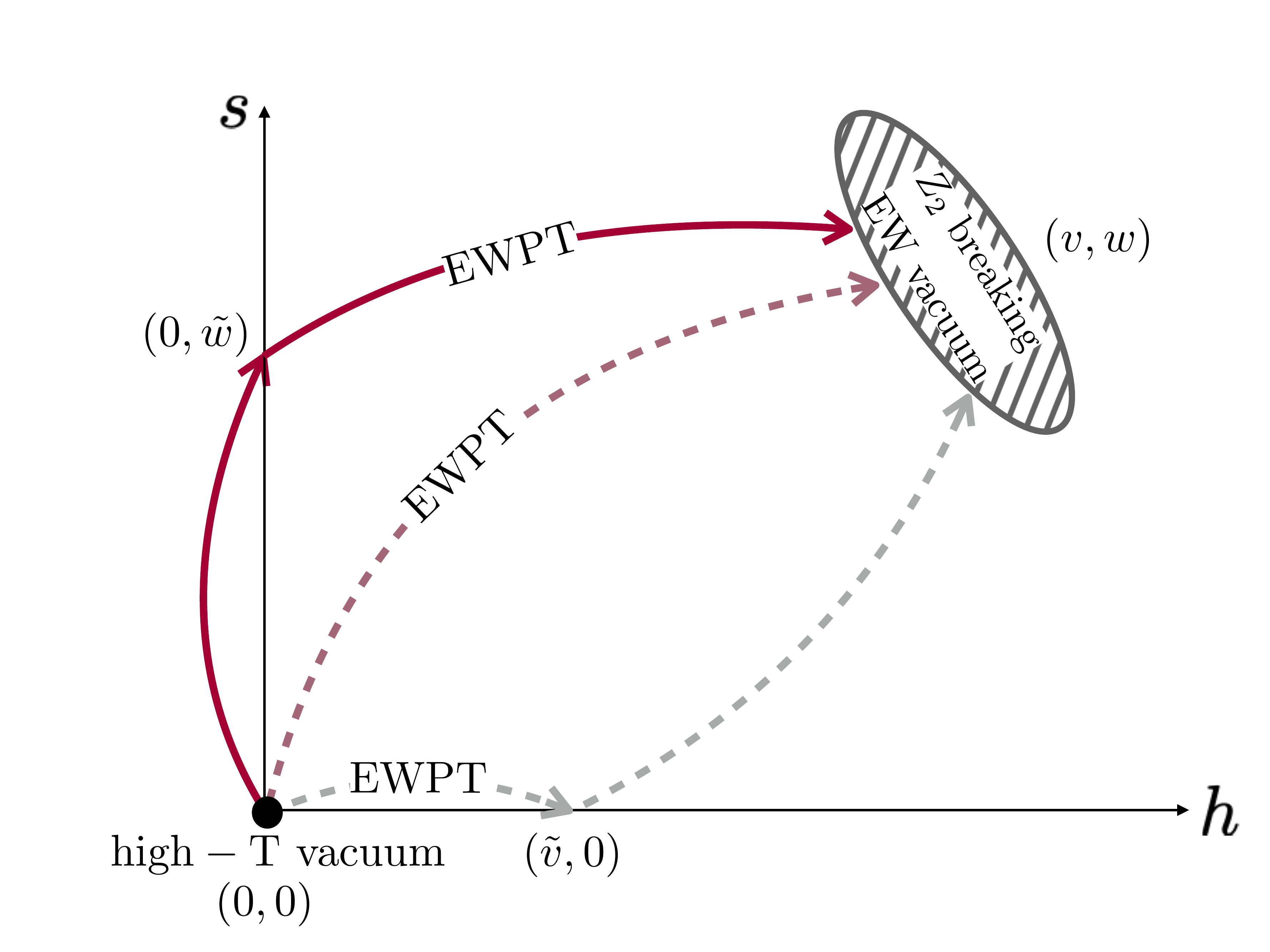}
         \caption{$Z_2$ spontaneous breaking scenarios with symmetry restoration.}
         \label{fig:EWPhT_patterns1}
     \end{subfigure}
     \begin{subfigure}[t]{0.32\textwidth}
        % \centering
         \includegraphics[width=\textwidth]{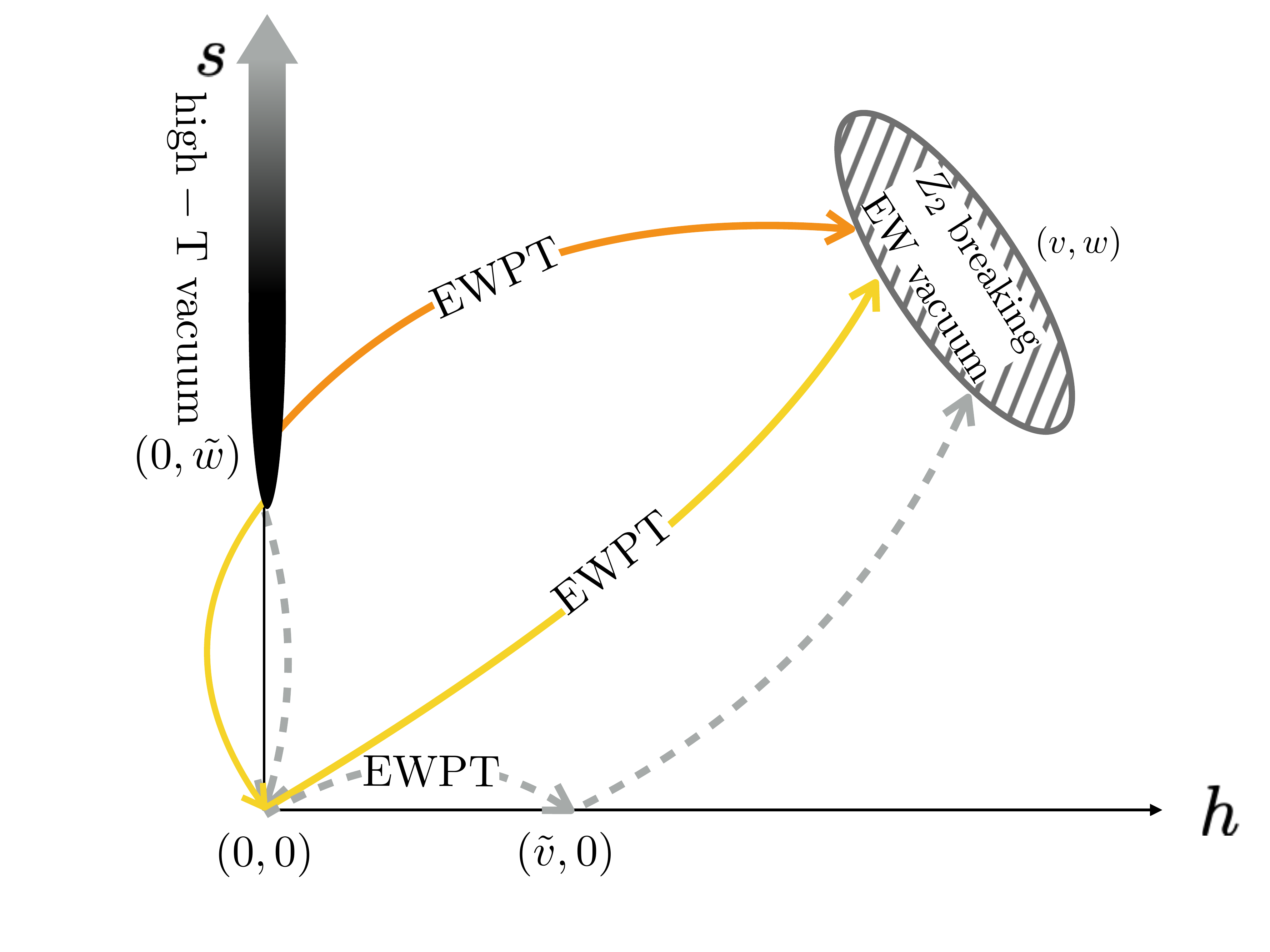}
         \caption{$Z_2$ spontaneous breaking scenarios with symmetry non-restoration.}
         \label{fig:EWPhT_patterns2}
     \end{subfigure}
     \begin{subfigure}[t]{0.32\textwidth}
       %  \centering
         \includegraphics[width=\textwidth]{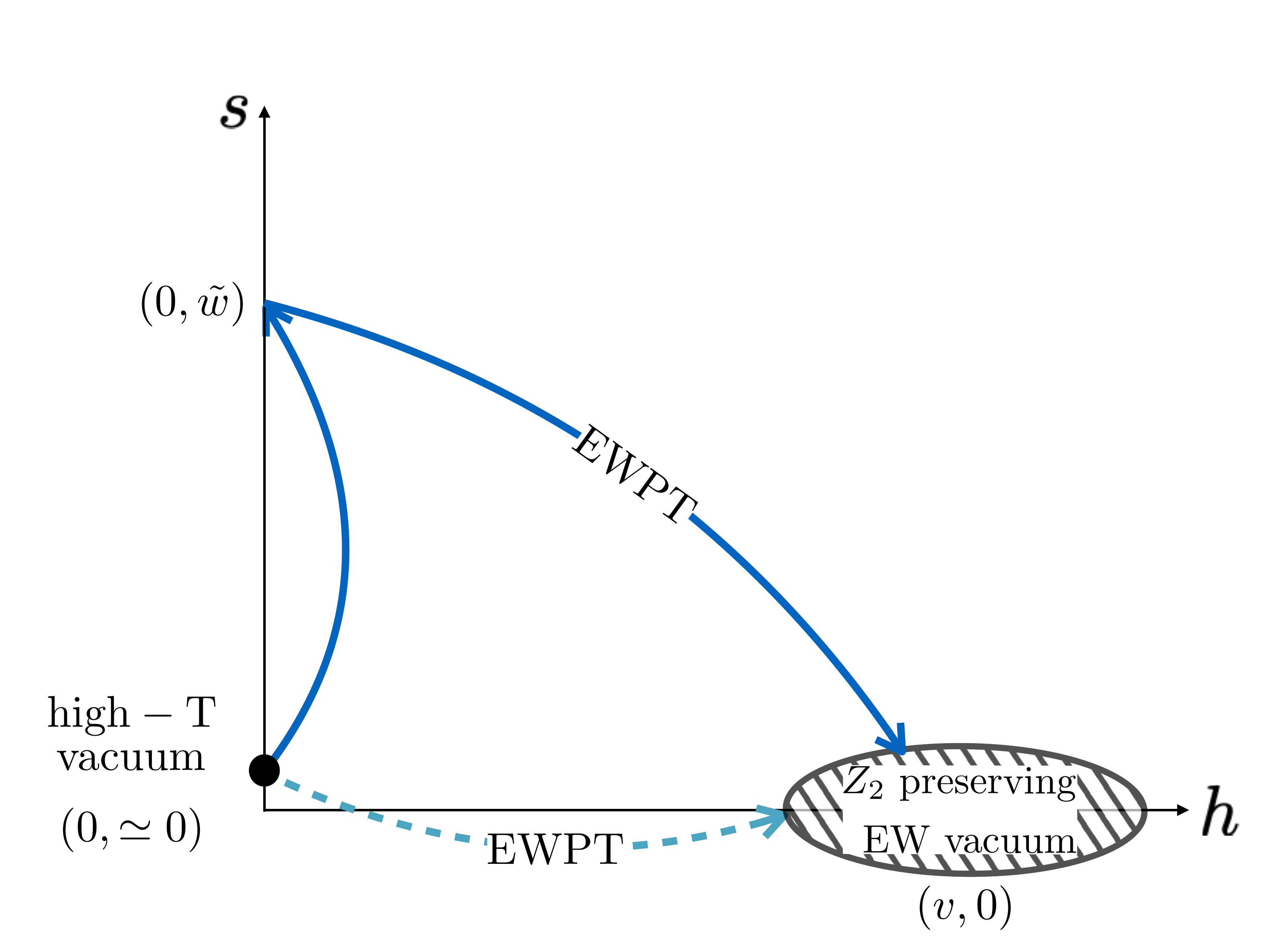}
         \caption{$Z_2$ preserving or explicit breaking scenarios.}
         \label{fig:EWPhT_pattern3}
     \end{subfigure}
        \caption{
        Schematics of EWPT patterns in the singlet extended SM in different scenarios relevant for Higgs exotic decays \cite{Kozaczuk:2019pet, Carena:2019une}. Solid lines correspond to patterns where EWPT can be strongly first-order, while dashed lines correspond to patterns where EWPT is weakly first-order, second order or a crossover. Transition step in which the electroweak symmetry is first broken, thus relevant for baryogenesis, is labeled as ``EWPT''.
        % The patterns of first-order EWPT in the singlet extended SM with a spontaneous breaking $Z_2$, taken from Ref.~\cite{Carena:2019une}. Left: $Z_2$ restoration scenarios. Right: $Z_2$ non-restoration scenarios. See the text for details.
        % \ZL{Yikun, please adjust the placement of the figures so they have equal height.}
        }
        \label{fig:EWPhT_patterns}
\end{figure}

In the viable light scalar scenarios that yield strongly first-order EWPTs, symmetry breaking in the thermal history generically proceeds through a two-step transition, while a one-step transition is feasible in limited regions of parameter space.
The step where the electroweak symmetry is first broken is required to be strongly first-order. The possibilities are illustrated in Fig.~\ref{fig:EWPhT_patterns}
\footnote{In the general non-symmetric scenario, one is free to shift the singlet by a constant without changing the physical predictions of the theory. This shift is often used to either remove the tadpole term in the potential or set the singlet vacuum expectation value to zero. We choose the latter.}.
Since the SM phase transition is a crossover, if the singlet $s$ is to catalyze a strongly first-order phase transition, we expect its couplings to the Higgs cannot be too small in order to induce a sufficiently large deformation to the scalar potential in the early universe. Thus we generally expect that the exotic Higgs branching ratio into new scalars compatible with first-order phase transitions is bounded from below. The specific implications for the surviving light scalar parameter space depend on the (non-)realization of the $Z_2$ symmetry, as we now discuss.
% I left some other minor comments, should we address them now? 
% give me 5min and I'll be back ready for zoom
%sure, should we skype or zoom maybe? :) great

In the $Z_2$-preserving case, a sizeable quartic coupling $s^2|H|^2$ is crucial in generating a two-step phase transition (with the second step a first-order EWPT).  
In the general $Z_2$-explicit breaking scenario, the singlet cubic term $s^3$ can also be important in providing the tree-level barrier for a first-order EWPT. However, in the spontaneously-broken $Z_2$ case, at tree-level, the singlet $s^3$ term is absent, while the only allowed portal coupling, i.e., the quartic term $s^2|H|^2$, is related to the scalar-Higgs mixing angle and thus stringently constrained by Higgs precision tests. 
% In addition, even after the $Z_2$ is spontaneously broken, the $|H|^{2n}$ operators can only be generated at loop level with a suppressed contribution to the tree-level potential barrier. Therefore, the parameter space where the spontaneous breaking $Z_2$ scenario can trigger a first-order EWPT is not a priori clear.

% The EWPT in the singlet extended SM with a spontaneous breaking $Z_2$ is studied in detail in Ref.~\cite{Carena:2019une}. 
% The authors
However Ref.~\cite{Carena:2019une} demonstrated that the joint scalar sector with a spontanously-broken $Z_2$ can indeed realize a strong first-order EWPT while satisfying the current constraints on Higgs properties.  This analysis includes the
% \JS{to do: fine-tune specifics of ``complete''} {\color{blue} \sout{ complete one-loop zero-temperature and thermal corrections}} 
full one-loop radiative corrections from zero-temperature and thermal contributions, as well as the daisy resummation.
The surviving parameter space exhibits a sharp correlation between the first-order EWPT and the Higgs exotic decay. The region that realizes a strongly first-order EWPT parameter space while remaining compatible with Higgs properties features $m_s\lesssim25$ GeV and gives exotic decay branching ratios ${\rm Br}(h\to ss)\sim10^{-6}-10^{-1}$.
% , as shown in the light brown region in~Fig.~\ref{fig:bounds_2}.  
We extract from~Ref.~\cite{Carena:2019une} the motivated region in such a scenario that is also compatible with the current precision measurement constraints, as shown by the brown shadowed region in~Fig.~\ref{fig:bounds_2}. A more intense scan might reveal more tuned parameters, beyond the large scans done in~Ref.~\cite{Carena:2019une}, and cover further regions of smaller Higgs mass and lower branching ratio.

In the general non-symmetric  scenario, Ref.~\cite{Kozaczuk:2019pet}
shows that requiring a strong first-order EWPT together with a small amount of mixing between the Higgs and the new scalar implies a lower bound on the magnitude of the $hss$ coupling, and therefore a lower bound on the Higgs branching ratio for $h \to ss$. Ref.~\cite{Kozaczuk:2019pet} considers the leading contributions to the finite-temperature effective potential, keeping terms up to $\mathcal{O}(g^3)$ in Landau gauge, and performs numerical scans for fixed values of $\cos\theta$ that are well below current experimental bounds. The region of parameter space compatible with first-order phase transitions at fixed $\sin\theta = 0.01$ is shown as a blue shaded region in~Fig.~\ref{fig:bounds_2}. 
In particular, global fits to Higgs properties require $m_s \lesssim 28$ GeV.
Semi-analytical arguments in the small mixing limit, based on an expansion of the finite temperature effective potential that retains terms up to $\mathcal{O}(g^2)$, can be combined to obtain a simple expression for a lower bound on the value of Br($h\to ss$), as a function of $m_s$, that is compatible with first-order phase transitions.  

Unlike Ref.~\cite{Carena:2019une}, Ref.~\cite{Kozaczuk:2019pet} restricts attention to the region of parameter space with $m_s \gtrsim 5$ GeV. This restriction is technical: the region of parameter space realizing first-order phase transitions certainly extends to smaller scalar masses, but the power-counting arguments that justify the approximations made in the finite-temperature potential begin to break down at low scalar masses where smaller values of the mixed quartic coupling must be considered.

Ref.~\cite{Kozaczuk:2019pet} also performed a similar analysis for the $Z_2$-symmetric scenario.  In this scenario, purely analytic arguments (again based on an $\mathcal{O}(g^2)$ approximation to the finite-temperature effective potential) can provide a lower bound on the invisible Higgs branching ratio that is in excellent agreement with the results of numerical scans. The identified surviving parameter space of interest is shown in Fig.~4 in Ref.~\cite{Kozaczuk:2019pet}.  In this case direct LHC searches for $h\to$ invisible requires $m_s \lesssim 20$ GeV.  %\JS{this can prob be updated}\ZL{Just checked the 2021 CMS results, it is still 18\%. So I think we don't need to update here.}

In all three scenarios, the parameter space compatible with both exotic Higgs decays and a first-order EWPT predicts exotic branching ratios Br($h\to ss$) that are bounded from below. The branching ratios of interest are in the range $10^{-6}-10^{-1}$, with smaller branching ratios possible at smaller values of $m_s$. The resulting parameter space poses an attractive target for both visible and invisible exotic Higgs decay searches at current and future colliders, as discussed in the next section. 

In summary, both studies suggest that searches for exotic Higgs decays at the LHC and future colliders can play a vital role in probing the nature of the EWPT in models with light scalars. Next-generation experiments are likely to either unearth evidence for or concretely rule out this class of scenarios.

\section{Current status and recent results in exotic decays}
%{\flushleft  {\color{blue} [Ke-Pan, Zhen, Tong]}}

The studies discussed in the previous section point us toward an intriguing, accessible signal in exotic Higgs  decays.
The 125 GeV SM-like Higgs boson can decay to pairs of new particles via the portal couplings. As the SM Higgs boson has a very narrow decay width $\Gamma_h=4.07$ MeV, even very tiny BSM couplings can have appreciable impacts on the decay branching ratios, making exotic Higgs decays a potent probe of beyond-the-SM interactions~\cite{Curtin:2013fra,Cepeda:2021rql}. This section summarizes the current status and future prospects for Higgs exotic decays that are relevant for the EWPT-motivated $h\to ss$ decays.

\begin{figure}[t]
\centering
\includegraphics[scale=0.33]{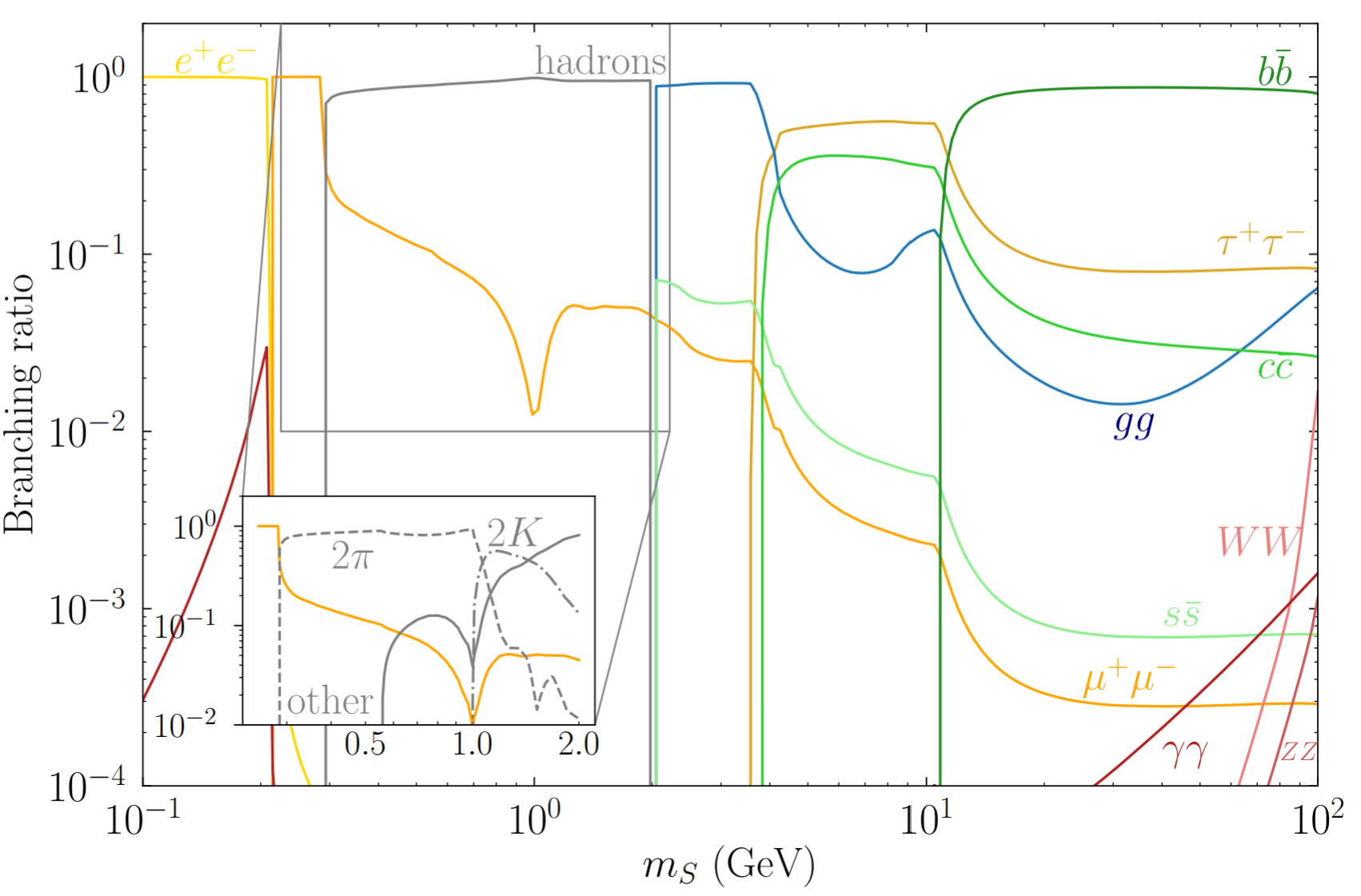}
\caption{Scalar singlet $s$ branching fractions mediated through mixing with the Higgs boson, taken from Ref.~\cite{Gershtein:2020mwi}.}
\label{fig:br}
\end{figure}

Current global fits constrain the Higgs exotic decay branching ratio to be $\leq16\%$ at 95\% C.L.~\cite{ATLAS:2021vrm}. The $h\to ss$ decay mentioned in the last section can lead to various final states according to the subsequent decay channels of the light scalar state $s$. In these SM plus singlet models, the $s$ decay is controlled by the $s$-$h$ mixing and inherits the Higgs-like hierarchical branching fractions following the corresponding fermion masses. The final state is dominated by $h\to ss\to bbbb$ for $m_s>2m_b\sim10$ GeV, and by $jjjj,jj\tau\tau$, and $\tau\tau\tau\tau$ for $m_s<10$ GeV. In Fig.~\ref{fig:br} we show the singlet decay branching fractions to various final states from Ref.~\cite{Gershtein:2020mwi}, building on the work of Refs.~\cite{Spira:1997dg,Winkler:2018qyg}. In general, the final state arising from $h\to ss$ can be written as $XXYY$, where $X$ and $Y$ represent (possibly) different particles. Beyond these visible decays, if $s$ decays dominantly to dark particles or is stable on collider time scales, the signal would be invisible Higgs decay. 
For instance, the $Z_2$ symmetry in the $s\to -s$ could be exact, and then the scalar $s$ could be stable and hence appear as missing energy (prospects for this case were surveyed in \cite{Kozaczuk:2019pet}). Other generalization of $Z_2$ symmetric SM-singlet extensions could further alter the cosmological history, even achieving electroweak non-restoration, and leading to Higgs invisible decay phenomenology~\cite{Carena:2021onl}.
Moreover, supersymmetric extensions of the Higgs potential, as for example in the  next-to-minimal-supersymmetric-standard-model (NMSSM) can give raise to strongly first order phase transition and induce interesting phenomenology that deserves a detailed study~\cite{Baum:2020vfl}. 
%Regarding the invisible signatures, we will discuss at the end of this section.

\begin{table}[t]\footnotesize\renewcommand\arraystretch{1.5}\centering
\begin{tabular}{|c|c|c|c|c|c|c|c|c|c|c|c|c|c|c|c|c|c|c|c|c|c|} \hline
Final state & Production mode & $m_s$ range [GeV] & $\mathcal{L}$ [fb$^{-1}$] & Collaboration \\ \hline
\multirow{3}{*}{$\mu\mu\mu\mu$} & \multirow{3}{*}{$gg$ fusion} & $[0.25,8.5]$ & 35.9 & CMS~\cite{CMS:2018jid}  \\ \cline{3-5}
 & & $[4,8]\cup[11.5,60]$ & 137 & CMS~\cite{CMS:2021pcy}  \\ \cline{3-5}
 &  & $[1.2,2]\cup[4.4,8]\cup[12,60]$ & 139 & ATLAS~\cite{ATLAS:2021ldb}  \\ \hline
\multirow{2}{*}{$\mu\mu\tau\tau$} & \multirow{2}{*}{$gg$ fusion} & $[3.6,21]$ & 35.9 & CMS~\cite{CMS:2020ffa} \\ \cline{3-5}
&  & $[15,62.5]$ & 35.9 & CMS~\cite{CMS:2018qvj} \\ \hline
$\mu\mu\tau\tau$ & $gg$ fusion & $[4,15]$ & 35.9 & CMS~\cite{CMS:2019spf} \\ \hline
\multirow{2}{*}{$bb\mu\mu$} & \multirow{2}{*}{$gg$ fusion} & $[18,60]$ & 139 & ATLAS~\cite{ATLAS:2021hbr} \\ \cline{3-5}
&  & $[20,62.5]$ & 35.9 & CMS~\cite{CMS:2018nsh} \\ \hline
$bb\tau\tau$ & $gg$ fusion & $[15,60]$ & 35.9 & CMS~\cite{CMS:2018zvv} \\ \hline
\multirow{2}{*}{$bbbb$} & $Zh$ & $[15,30]$ & 36.1 & ATLAS~\cite{ATLAS:2018pvw} \\ \cline{2-5}
& $Wh/Zh$ & $[20,60]$ & 36.1 & ATLAS~\cite{ATLAS:2020ahi} \\ \hline
$\gamma\gamma\gamma\gamma$ & $gg$ fusion & $[15,60]$ & 132 & CMS~\cite{CMS:2021bvh} \\ \hline
$\gamma\gamma jj$ & VBF & $[20,60]$ & 36.7 & ATLAS~\cite{ATLAS:2018jnf} \\ \hline
\end{tabular}
\caption{The existing experimental searches for exotic Higgs decays $h\to ss\to XXYY$ at the 13 TeV LHC. Modified from Table 1 of Ref.~\cite{Cepeda:2021rql}.}\label{table:XXYY}
\end{table}

%Below we introduce the possible final states and current experimental limits. 
%The LHC discovered the Higgs boson and continues to reveal exciting physics opportunities from the Higgs exotic decay program. 
We summarize existing searches at the LHC for $h\to ss\to XXYY$ final states at the 13 TeV LHC in Table~\ref{table:XXYY}. This table lists the final state targeted by a given analysis, together with the utilized production mode and the corresponding integrated luminosity, as well as the intermediate scalar $s$ mass range considered in each search. Except for the $bbbb$ final states, searches for all the other final states at the LHC involve at least a pair of non-hadronic final states, such as photons, muons, or tau leptons. 
%Background considerations drive these search channels at the LHC.

\begin{figure}
\centering
\includegraphics[scale=0.49]{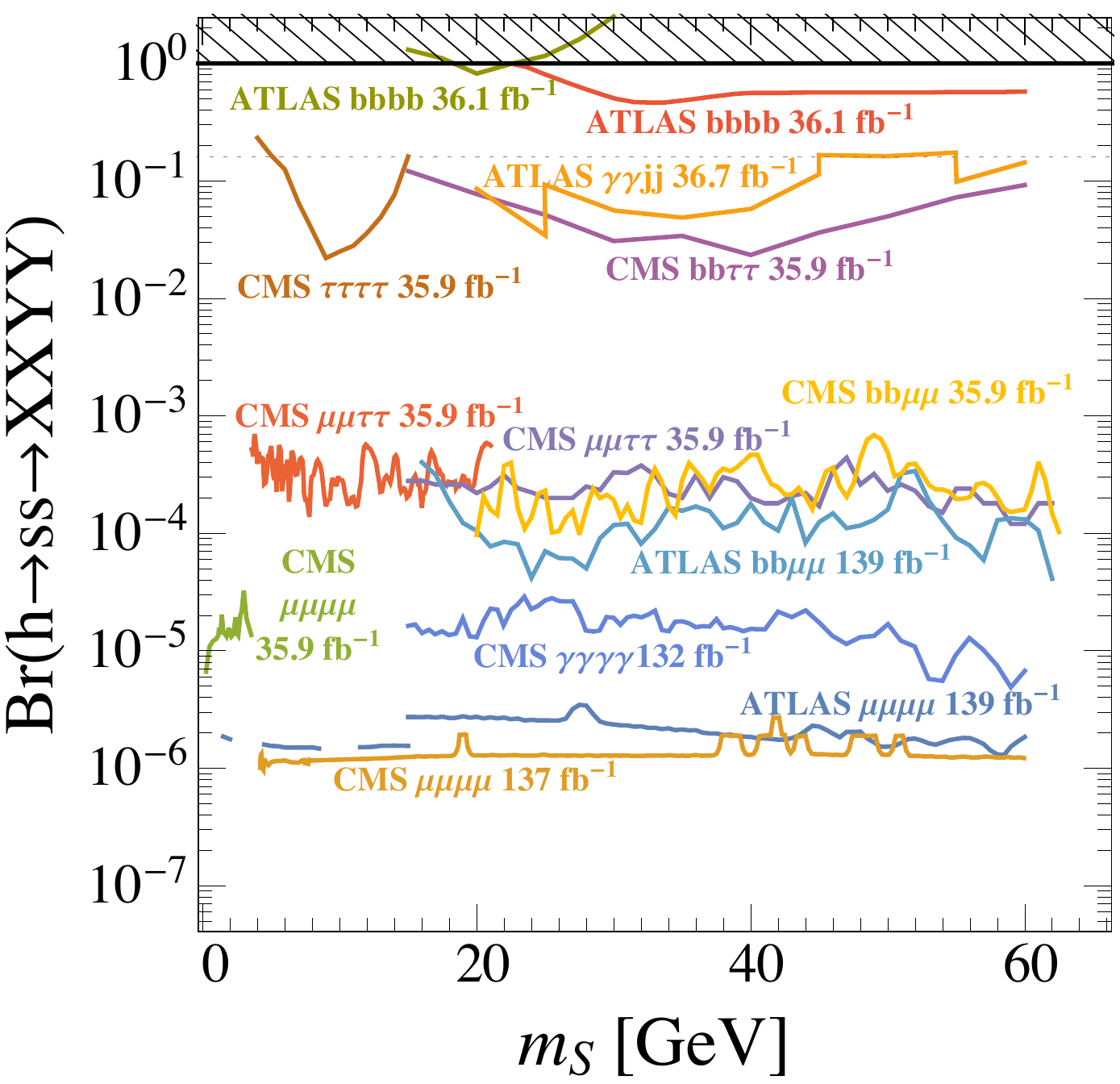}\qquad
\includegraphics[scale=0.49]{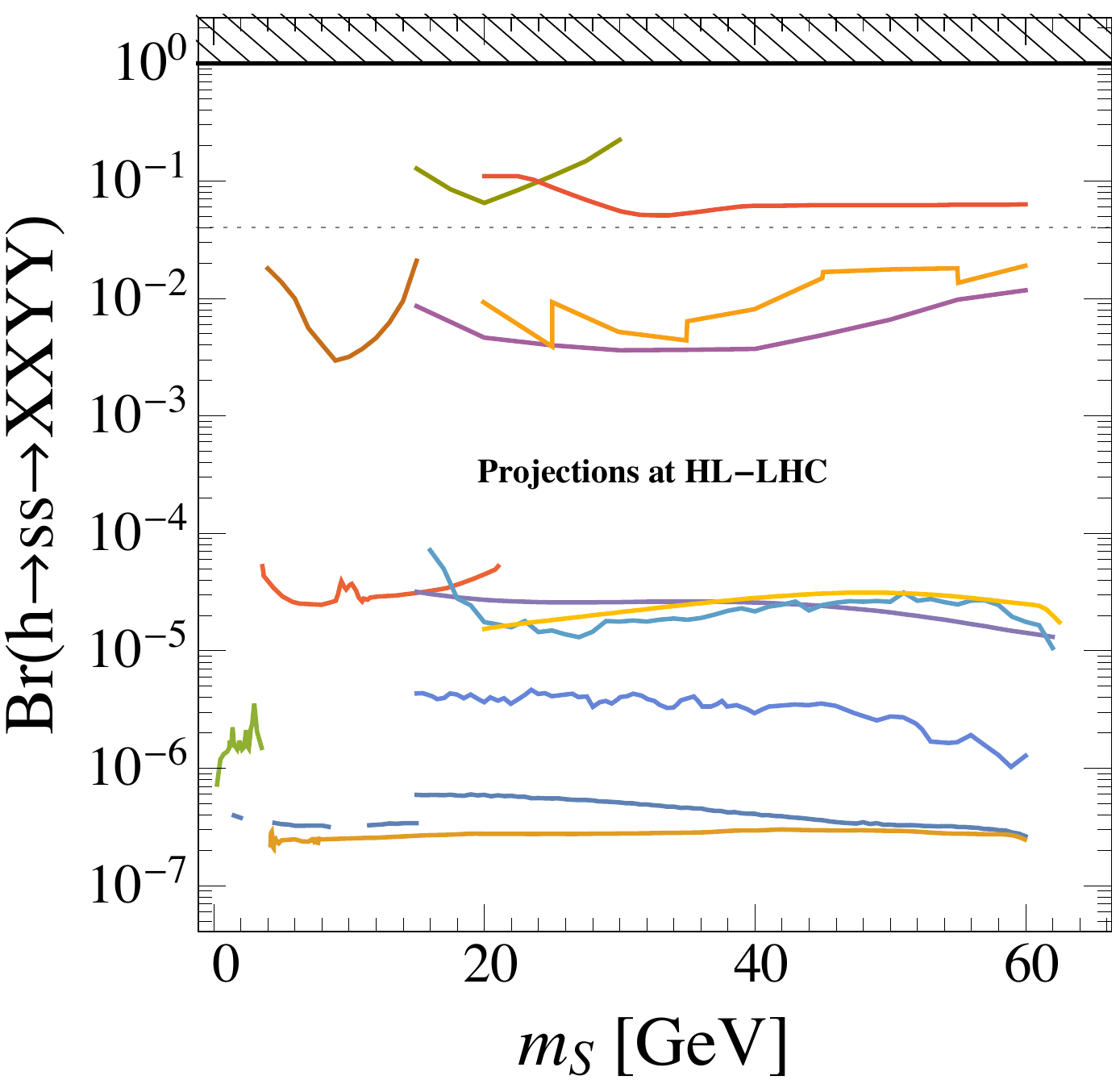}
\caption{Current bounds (left panel) on exotic Higgs  decays $h\to ss\to XXYY$ and corresponding projections (right panel) at the HL-LHC.
The 
%upper and lower 
horizontal dotted line is the current and future projection of upper limit for the exotic Higgs branching ratio from global fits to Higgs properties (16\% and 4\% respectively).
%and the statistical limit given $10^6$ Higgs at future lepton colliders, respectively. %The brown and light blue shadowed regions are the strong first-order EWPT regions identified in Refs.~\cite{Kozaczuk:2019pet,Carena:2019une} as described in the text.
}
\label{fig:bounds}
\end{figure}

In Fig.~\ref{fig:bounds} we show  current LHC constraints and future HL-LHC projections on the branching fraction of $H\to XXYY$ final states as a function of the intermediate scalar mass. The HL-LHC projections are derived using the simple assumption that all uncertainties can be taken to scale as $1/\sqrt{L}$.
%a simple rescaling of the statistical uncertainties with luminosity \JS{how were systematic uncertainties handled?}. 
Searches in these individual final states exclude regions above the lines. We can see that the $\mu\mu\mu\mu$ channel provides a strong limit on Br($h\to ss\to XXYY$) to around $10^{-6}$-$10^{-5}$ across the scalar mass. The $\gamma\gamma\gamma\gamma$ channel also makes a stringent $\sim10^{-5}$ bound. The constraints from $bb\mu\mu$ and $\mu\mu\tau\tau$ channels are a bit weaker, around $10^{-4}-10^{-3}$, but still stronger than the $bb\tau\tau$, $\tau\tau\tau\tau$ and $\gamma\gamma jj$ bounds which are around $10^{-2}-10^{-1}$. The current $bbbb$ bounds are typically higher than the allowed maximal exotic branching ratio (16\%), but the HL-LHC projections can reach a few percent. On the other hand, the $\mu\mu\mu\mu$ channel can touch $10^{-7}$ at the HL-LHC.

\begin{figure}
\centering
\includegraphics[scale=0.49]{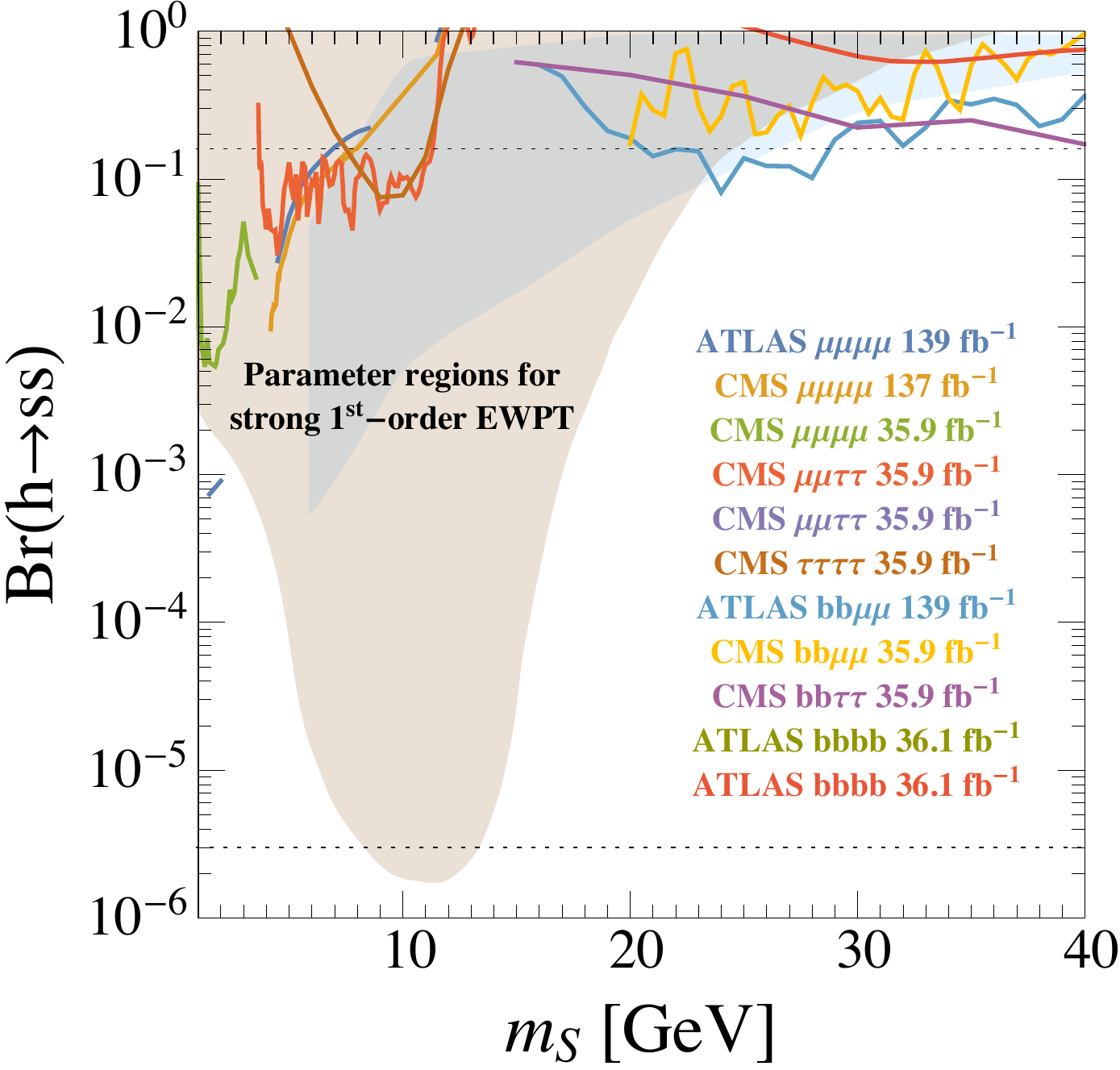}\qquad
\includegraphics[scale=0.49]{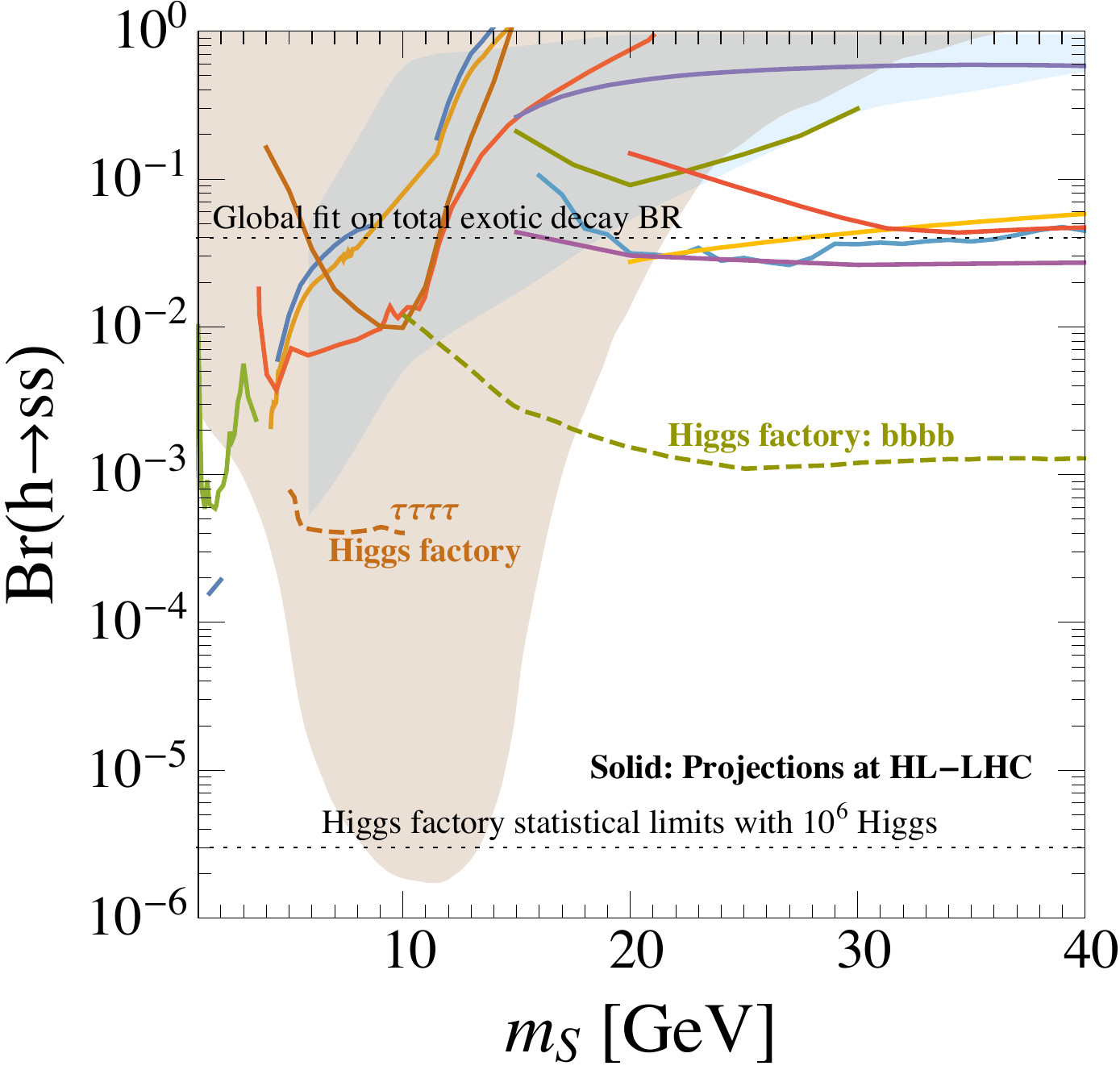}
\caption{The current bounds on Higgs exotic decay $h\to ss$ and the projections at the HL-LHC, assuming the $s$ decays to SM particles are mediated by the mixing, and the corresponding branching ratios are taken from Ref.~\cite{Gershtein:2020mwi}. The upper and lower horizontal dotted lines are the expected upper limit for Higgs exotic decay branching ratio at the HL-LHC (4\%~\cite{deBlas:2019rxi}) and statistical limit of $10^6$ Higgs at future lepton colliders, respectively. The brown and light blue shadowed regions are the strong first-order EWPT regions from Refs.~\cite{Kozaczuk:2019pet,Carena:2019une}, see text for details. Projections of the reach of future lepton colliders are shown in dashed lines.}
\label{fig:bounds_2}
\end{figure}

The bounds on Br($h\to ss$) can be derived from those on Br($h\to ss\to XXYY$) once the $s\to XX/YY$ branching ratios are given. Assuming the $s$ decay branching ratios are dominated by the $h$-$s$ mixing (see Fig.~\ref{fig:br}), the bounds on Br($h\to ss$) are given in Fig.~\ref{fig:bounds_2}. We can see that the hierarchies of various channels are significantly affected compared to those in Fig.~\ref{fig:bounds}. For $m_s\lesssim 10$ GeV, the strongest bounds are still from the $\mu\mu$-relevant channels, e.g. $\mu\mu\mu\mu$ for $m_s\lesssim3.5$ GeV and $\mu\mu\tau\tau$ for $3.5~{\rm GeV}\lesssim m_s\lesssim 10$ GeV, respectively. For $m_s\gtrsim10$ GeV, $bb$ is the main decay channel of $s$, making $bb$-relevant channels most sensitive. As a result, the most stringent bounds for $10~{\rm GeV}\lesssim m_s<62.5$ GeV is $bb\mu\mu$ and $bb\tau\tau$.

In Fig.~\ref{fig:bounds_2} we show the projected reach of the $\sim240$ GeV $e^+e^-$ colliders with an integrated luminosity of $5~{\rm ab}^{-1}$ for the $\tau\tau\tau\tau$~\cite{Shelton:2021xwo} and $bbbb$~\cite{Liu:2016zki} channels. The projections for $qqqq/gggg$ and $cccc$ channels can be found in Ref.~\cite{Liu:2016zki}, which we do not show here. A ILC-based simulation (250 GeV, 0.9 ab$^{-1}$) for the $bbbb$ channel is done by Ref.~\cite{Kato:2021ILC4b} and find similar projections. There are room for further improvement, e.g., usings Machine Learning to deal with complex signals and backgrounds for Higgs exotic decays~\cite{Jung:2021tym}.

The strongly first-order EWPT parameter space for the spontaneous $Z_2$ breaking model~\cite{Carena:2019une} and general singlet scalar extension of the SM with mixing angle $\sin\theta=0.01$~\cite{Kozaczuk:2019pet} are indicated by the brown and light blue shaded regions in the Br($h\to ss$)-$m_s$ plane in Fig.~\ref{fig:bounds_2}, respectively. Combining with the current bounds at the LHC, we see that the exotic Higgs decay searches have already probed a visible fraction of the EWPT parameter space, especially for the low mass region $m_s\lesssim10$ GeV. For the high mass region $m_s>10$ GeV, the direct constraints from the $bb$-relevant channels are slightly weaker than the indirect bound (16\%~\cite{ATLAS-CONF-2021-053}) from the exotic Higgs  decay. At the HL-LHC, the Br($h\to ss$) reach in both the low and high mass regions are significantly improved, while the expected reach from direct searches at high masses is still comparable with the expected indirect bounds (4\%~\cite{deBlas:2019rxi}).  Future Higgs factories can greatly improve the coverage of EWPT parameter space, as shown in the dashed lines in the right panel of Fig.~\ref{fig:bounds_2}. Combining the $\tau\tau\tau\tau$ and $bbbb$ channels, future Higgs factories will be able to cover almost all the EWPT parameter space for the general singlet scalar extension SM with mixing angle $\sin\theta=0.01$~\cite{Kozaczuk:2019pet}. 
%On the other hand, in the spontaneous $Z_2$ breaking case~\cite{Carena:2019une}, there is some EWPT parameter space with too low Br($h\to ss$) branching ratio (some even lower than the 
For tiny mixing angles $\sin\theta$, that  render the singlet scalar long-lived, recent studies ~\cite{Feng:2022inv} consider a class of dark Higgs searches at  various intensity frontier machines that can provide useful probes.
Note that the statistical limit of a typical Higgs factory with $10^6$ Higgs is $3\times10^{-6}$. Given the clean environment with low background at lepton colliders, we conclude that there is ample scope for future studies in exotic decays.

%\underline{$jjjj$}. If $m_s\lesssim10$ GeV, $s$ can only decay to light flavor jets. While no specific experimental search is done for this channel, there are a few LHC phenomenological studies available~\cite{Falkowski:2010hi,Chen:2010wk,Lewis:2012pf}, which make use of the jet substructure techniques to search for $h\to ss\to c\bar cc\bar c$ or $gggg$ cascade decays where the decay products of each $s$ are collected into a fat-jet.

%\underline{Higgs invisible decay}. There have been experimental searches at the 13 TeV LHC~\cite{ATLAS:2019cid,ATLAS:2021gcn,CMS:2022qva}

\section{Summary and Outlook}
%\ZL{Will give it a try.}
Determining the true nature of the electroweak phase transition is a fundamental question in particle physics, and  connects searches at energy frontier facilities to the history of the early universe.  While much study has been devoted to discovery prospects for theories where heavy ($m>m_h$) new physics drives the electroweak phase transition strongly first-order, we emphasize that there  exists an experiemntally viable region of parameter space where {\em light} new states can accomplish the same goal.  Following the studies in \cite{Kozaczuk:2019pet,Carena:2019une}, we describe the frameworks where the simple singlet scalar extension of the SM can realize a strongly first-order EWPT along with allowing for exotic Higgs decays, $h\to ss$. Such new physics tragets are 
 an interesting and accessible target for both the LHC and future Higgs factories.  

%We show that the consideration of enhancing electroweak phase transition via the simplest singlet extension leads to an intriguing physics opportunity of Higgs exotic decay. From three scenarios, the spontaneous $Z_2$ breaking, explicit $Z_2$ breaking, and $Z_2$ preserving, as discussed in \autoref{sec:theory} and depicted in \autoref{fig:EWPhT_patterns}, they all point towards accessible Higgs exotic decay signals at current and future colliders. The vibrant search program at the LHC is already probing the EWPT-motivated parameter space; These Higgs exotic decay searches can be translated into the model parameter space, and 

At the LHC we find that direct searches in the $bbbb$, $bb\mu\mu$ and $\mu\mu\mu\mu$ final states lead the sensitivities that can cover new singlet masses below 10--15 GeV with Higgs exotic branching ratios as low as  10$^{-2}$, as  shown in the left panel of~Fig.~\ref{fig:bounds_2}. The full run of the HL-LHC and future lepton collider Higgs factories can access most of the relevant parameter space for singlet SM extension models that enable a  strongly first  order EWPT, with the current projections  shown in the right panel of~Fig.~\ref{fig:bounds_2}. The $bbbb$ and $\tau\tau\tau\tau$ channels have the best sensitivities at future lepton colliders.  The sensitivity projections shown  are based on studies of the signal $e^{+}e^-\to hZ$ in various exclusive final states and assuming leptonic decays of the $Z$. There is ample room for improvement by including more final states for the Higgs decay, especially in the low mass regions, and by considering hadronic $Z$ decays. Altogether a combination of future lepton and hadron colliders  offer excellent prospects for full coverage of the strongly first order EWPT-motivated region of relevance for electroweak baryogenesis models.
%eventually fully probing the well-motivated parameter regions.

\acknowledgements{This
manuscript has been authored by Fermi Research Alliance, LLC under Contract No. DE-AC02-07CH11359 with the U.S. Department of Energy, Office of Science, Office of High Energy Physics. Z.L. is supported in part by the U.S.~Department of Energy (DOE) under grant No. DE-SC0022345. J.S.~is supported in part by U.S.~DOE CAREER grant DE-SC0017840. K.-P.X. is supported by the University of Nebraska-Lincoln.
MJRM was supported in part under U.S.~DOE Contract No. DE-SC0011095 and National Natural Science Foundation of China Grant No. 19Z103010239. Y.W.~is supported by the Walter Burke Institute for Theoretical Physics. T.O. is supported by the Visiting Scholars Program of Universities Research Association.}

%\clearpage
% \section{Objectives}

% In light of these results, we propose a Snowmass study on the implications of exotic Higgs decay searches at colliders for the early Universe. We envision the objectives of such a study to include: 
% \begin{itemize}
% \item Highlighting and quantifying the extent to which exotic Higgs decay measurements at the LHC and future colliders are sensitive to a first-order EWPT catalyzed by light BSM degrees of freedom.  Both scenarios with and without a $Z_2$ symmetry are of interest.

% %\item Extending the results of Refs.~\cite{Kozaczuk:2019pet, Carena:2019une} to cover additional parameter space. Of particular interest are low masses where additional search strategies (e.g. direct $s$ production in heavy meson decays) can become relevant.

% \item Identifying opportunities for future study that will enable current and next generation of collider experiments to harness the full power of the precision Higgs program in exploring the nature of the EWPT.
% \end{itemize}

%We welcome the participation of other members of the community who may be interested to join us in this study.

\bibliographystyle{utphys}
\bibliography{Higgs_Decays_EWPT}

\providecommand{\href}[2]{#2}\begingroup\raggedright\begin{thebibliography}{10}

\bibitem{Morrissey:2012db}
D.~E. Morrissey and M.~J. Ramsey-Musolf, ``{Electroweak baryogenesis},''
  \href{http://dx.doi.org/10.1088/1367-2630/14/12/125003}{{\em New J. Phys.}
  {\bf 14} (2012)  125003},
\href{http://arxiv.org/abs/1206.2942}{{\tt arXiv:1206.2942 [hep-ph]}}.
%%CITATION = ARXIV:1206.2942;%%.

\bibitem{Caprini:2015zlo}
C.~Caprini {\em et al.}, ``{Science with the space-based interferometer eLISA.
  II: Gravitational waves from cosmological phase transitions},''
  \href{http://dx.doi.org/10.1088/1475-7516/2016/04/001}{{\em JCAP} {\bf 1604}
  (2016) no.~04, 001},
\href{http://arxiv.org/abs/1512.06239}{{\tt arXiv:1512.06239 [astro-ph.CO]}}.
%%CITATION = ARXIV:1512.06239;%%.

\bibitem{Caprini:2019egz}
C.~Caprini {\em et al.}, ``{Detecting gravitational waves from cosmological
  phase transitions with LISA: an update},''
\href{http://arxiv.org/abs/1910.13125}{{\tt arXiv:1910.13125 [astro-ph.CO]}}.
%%CITATION = ARXIV:1910.13125;%%.

\bibitem{Wainwright:2009mq}
C.~Wainwright and S.~Profumo, ``{The Impact of a strongly first-order phase
  transition on the abundance of thermal relics},''
  \href{http://dx.doi.org/10.1103/PhysRevD.80.103517}{{\em Phys. Rev. D} {\bf
  80} (2009)  103517}, \href{http://arxiv.org/abs/0909.1317}{{\tt
  arXiv:0909.1317 [hep-ph]}}.

\bibitem{Profumo:2007wc}
S.~Profumo, M.~J. Ramsey-Musolf, and G.~Shaughnessy, ``{Singlet Higgs
  phenomenology and the electroweak phase transition},''
  \href{http://dx.doi.org/10.1088/1126-6708/2007/08/010}{{\em JHEP} {\bf 08}
  (2007)  010},
\href{http://arxiv.org/abs/0705.2425}{{\tt arXiv:0705.2425 [hep-ph]}}.
%%CITATION = ARXIV:0705.2425;%%.

\bibitem{Kozaczuk:2019pet}
J.~Kozaczuk, M.~J. Ramsey-Musolf, and J.~Shelton, ``{Exotic Higgs boson decays
  and the electroweak phase transition},''
  \href{http://dx.doi.org/10.1103/PhysRevD.101.115035}{{\em Phys. Rev. D} {\bf
  101} (2020) no.~11, 115035}, \href{http://arxiv.org/abs/1911.10210}{{\tt
  arXiv:1911.10210 [hep-ph]}}.

\bibitem{Carena:2019une}
M.~Carena, Z.~Liu, and Y.~Wang, ``{Electroweak phase transition with
  spontaneous Z$_{2}$-breaking},''
  \href{http://dx.doi.org/10.1007/JHEP08(2020)107}{{\em JHEP} {\bf 08} (2020)
  107}, \href{http://arxiv.org/abs/1911.10206}{{\tt arXiv:1911.10206
  [hep-ph]}}.

\bibitem{Curtin:2013fra}
D.~Curtin {\em et al.}, ``{Exotic decays of the 125 GeV Higgs boson},''
  \href{http://dx.doi.org/10.1103/PhysRevD.90.075004}{{\em Phys. Rev. D} {\bf
  90} (2014) no.~7, 075004}, \href{http://arxiv.org/abs/1312.4992}{{\tt
  arXiv:1312.4992 [hep-ph]}}.

\bibitem{Abada:2019lih}
{\bf FCC} Collaboration, A.~Abada {\em et al.}, ``{FCC Physics
  Opportunities},''
\href{http://dx.doi.org/10.1140/epjc/s10052-019-6904-3}{{\em Eur. Phys. J.}
  {\bf C79} (2019) no.~6, 474}.
%%CITATION = EPHJA,C79,474;%%.

\bibitem{CEPCStudyGroup:2018ghi}
{\bf CEPC Study Group} Collaboration, M.~Dong {\em et al.}, ``{CEPC Conceptual
  Design Report: Volume 2 - Physics \& Detector},''
\href{http://arxiv.org/abs/1811.10545}{{\tt arXiv:1811.10545 [hep-ex]}}.
%%CITATION = ARXIV:1811.10545;%%.

\bibitem{Benedikt:2018csr}
{\bf FCC} Collaboration, A.~Abada {\em et al.}, ``{FCC-hh: The Hadron
  Collider}: {Future Circular Collider Conceptual Design Report Volume 3},''
  \href{http://dx.doi.org/10.1140/epjst/e2019-900087-0}{{\em Eur. Phys. J. ST}
  {\bf 228} (2019) no.~4, 755--1107}.

\bibitem{Abada:2019zxq}
{\bf FCC} Collaboration, A.~Abada {\em et al.}, ``{FCC-ee: The Lepton
  Collider}: {Future Circular Collider Conceptual Design Report Volume 2},''
  \href{http://dx.doi.org/10.1140/epjst/e2019-900045-4}{{\em Eur. Phys. J. ST}
  {\bf 228} (2019) no.~2, 261--623}.

\bibitem{Cepeda:2021rql}
M.~Cepeda, S.~Gori, V.~M. Outschoorn, and J.~Shelton, ``{Exotic Higgs
  Decays},'' \href{http://arxiv.org/abs/2111.12751}{{\tt arXiv:2111.12751
  [hep-ph]}}.

\bibitem{deFlorian:2016spz}
{\bf LHC Higgs Cross Section Working Group} Collaboration, D.~de~Florian {\em
  et al.}, ``{Handbook of LHC Higgs Cross Sections: 4. Deciphering the Nature
  of the Higgs Sector},'' \href{http://arxiv.org/abs/1610.07922}{{\tt
  arXiv:1610.07922 [hep-ph]}}.

\bibitem{Cepeda:2019klc}
{\bf HL/HE WG2 group} Collaboration, M.~Cepeda {\em et al.}, ``{Higgs Physics
  at the HL-LHC and HE-LHC},''
\href{http://arxiv.org/abs/1902.00134}{{\tt arXiv:1902.00134 [hep-ph]}}.
%%CITATION = ARXIV:1902.00134;%%.

\bibitem{Liu:2016zki}
Z.~Liu, L.-T. Wang, and H.~Zhang, ``{Exotic decays of the 125 GeV Higgs boson
  at future $e^+e^-$ lepton colliders},''
  \href{http://dx.doi.org/10.1088/1674-1137/41/6/063102}{{\em Chin. Phys.} {\bf
  C41} (2017) no.~6, 063102},
\href{http://arxiv.org/abs/1612.09284}{{\tt arXiv:1612.09284 [hep-ph]}}.
%%CITATION = ARXIV:1612.09284;%%.

\bibitem{Gershtein:2020mwi}
Y.~Gershtein, S.~Knapen, and D.~Redigolo, ``{Probing naturally light singlets
  with a displaced vertex trigger},''
  \href{http://dx.doi.org/10.1016/j.physletb.2021.136758}{{\em Phys. Lett. B}
  {\bf 823} (2021)  136758}, \href{http://arxiv.org/abs/2012.07864}{{\tt
  arXiv:2012.07864 [hep-ph]}}.

\bibitem{ATLAS:2021vrm}
{\bf ATLAS} Collaboration, ``{Combined measurements of Higgs boson production
  and decay using up to $139$ fb$^{-1}$ of proton-proton collision data at
  $\sqrt{s}= 13$ TeV collected with the ATLAS experiment },''.

\bibitem{Spira:1997dg}
M.~Spira, ``{QCD effects in Higgs physics},''
  \href{http://dx.doi.org/10.1002/(SICI)1521-3978(199804)46:3<203::AID-PROP203>3.0.CO;2-4}{{\em
  Fortsch. Phys.} {\bf 46} (1998)  203--284},
  \href{http://arxiv.org/abs/hep-ph/9705337}{{\tt arXiv:hep-ph/9705337}}.

\bibitem{Winkler:2018qyg}
M.~W. Winkler, ``{Decay and detection of a light scalar boson mixing with the
  Higgs boson},'' \href{http://dx.doi.org/10.1103/PhysRevD.99.015018}{{\em
  Phys. Rev. D} {\bf 99} (2019) no.~1, 015018},
  \href{http://arxiv.org/abs/1809.01876}{{\tt arXiv:1809.01876 [hep-ph]}}.

\bibitem{Carena:2021onl}
M.~Carena, C.~Krause, Z.~Liu, and Y.~Wang, ``{New approach to electroweak
  symmetry nonrestoration},''
  \href{http://dx.doi.org/10.1103/PhysRevD.104.055016}{{\em Phys. Rev. D} {\bf
  104} (2021) no.~5, 055016}, \href{http://arxiv.org/abs/2104.00638}{{\tt
  arXiv:2104.00638 [hep-ph]}}.

\bibitem{Baum:2020vfl}
S.~Baum, M.~Carena, N.~R. Shah, C.~E.~M. Wagner, and Y.~Wang, ``{Nucleation is
  more than critical}: {A case study of the electroweak phase transition in the
  NMSSM},'' \href{http://dx.doi.org/10.1007/JHEP03(2021)055}{{\em JHEP} {\bf
  03} (2021)  055}, \href{http://arxiv.org/abs/2009.10743}{{\tt
  arXiv:2009.10743 [hep-ph]}}.

\bibitem{CMS:2018jid}
{\bf CMS} Collaboration, A.~M. Sirunyan {\em et al.}, ``{A search for pair
  production of new light bosons decaying into muons in proton-proton
  collisions at 13 TeV},''
  \href{http://dx.doi.org/10.1016/j.physletb.2019.07.013}{{\em Phys. Lett. B}
  {\bf 796} (2019)  131--154}, \href{http://arxiv.org/abs/1812.00380}{{\tt
  arXiv:1812.00380 [hep-ex]}}.

\bibitem{CMS:2021pcy}
{\bf CMS} Collaboration, ``{Search for low-mass dilepton resonances in Higgs
  boson decays to four-lepton final states in proton-proton collisions at
  $\sqrt{s}$ =13 TeV},'' \href{http://arxiv.org/abs/2111.01299}{{\tt
  arXiv:2111.01299 [hep-ex]}}.

\bibitem{ATLAS:2021ldb}
{\bf ATLAS} Collaboration, G.~Aad {\em et al.}, ``{Search for Higgs bosons
  decaying into new spin-0 or spin-1 particles in four-lepton final states with
  the ATLAS detector with 139 fb$^{-1}$ of $pp$ collision data at $\sqrt{s}=13$
  TeV},'' \href{http://arxiv.org/abs/2110.13673}{{\tt arXiv:2110.13673
  [hep-ex]}}.

\bibitem{CMS:2020ffa}
{\bf CMS} Collaboration, A.~M. Sirunyan {\em et al.}, ``{Search for a light
  pseudoscalar Higgs boson in the boosted $\mu\mu\tau\tau$ final state in
  proton-proton collisions at $\sqrt{s}=$ 13 TeV},''
  \href{http://dx.doi.org/10.1007/JHEP08(2020)139}{{\em JHEP} {\bf 08} (2020)
  139}, \href{http://arxiv.org/abs/2005.08694}{{\tt arXiv:2005.08694
  [hep-ex]}}.

\bibitem{CMS:2018qvj}
{\bf CMS} Collaboration, A.~M. Sirunyan {\em et al.}, ``{Search for an exotic
  decay of the Higgs boson to a pair of light pseudoscalars in the final state
  of two muons and two $\tau$ leptons in proton-proton collisions at $
  \sqrt{s}=13 $ TeV},'' \href{http://dx.doi.org/10.1007/JHEP11(2018)018}{{\em
  JHEP} {\bf 11} (2018)  018}, \href{http://arxiv.org/abs/1805.04865}{{\tt
  arXiv:1805.04865 [hep-ex]}}.

\bibitem{CMS:2019spf}
{\bf CMS} Collaboration, A.~M. Sirunyan {\em et al.}, ``{Search for light
  pseudoscalar boson pairs produced from decays of the 125 GeV Higgs boson in
  final states with two muons and two nearby tracks in pp collisions at
  $\sqrt{s}=$ 13 TeV},''
  \href{http://dx.doi.org/10.1016/j.physletb.2019.135087}{{\em Phys. Lett. B}
  {\bf 800} (2020)  135087}, \href{http://arxiv.org/abs/1907.07235}{{\tt
  arXiv:1907.07235 [hep-ex]}}.

\bibitem{ATLAS:2021hbr}
{\bf ATLAS} Collaboration, G.~Aad {\em et al.}, ``{Search for Higgs boson
  decays into a pair of pseudoscalar particles in the $bb\mu\mu$ final state
  with the ATLAS detector in $pp$ collisions at $\sqrt s$=13\,\,TeV},''
  \href{http://dx.doi.org/10.1103/PhysRevD.105.012006}{{\em Phys. Rev. D} {\bf
  105} (2022) no.~1, 012006}, \href{http://arxiv.org/abs/2110.00313}{{\tt
  arXiv:2110.00313 [hep-ex]}}.

\bibitem{CMS:2018nsh}
{\bf CMS} Collaboration, A.~M. Sirunyan {\em et al.}, ``{Search for an exotic
  decay of the Higgs boson to a pair of light pseudoscalars in the final state
  with two muons and two b quarks in pp collisions at 13 TeV},''
  \href{http://dx.doi.org/10.1016/j.physletb.2019.06.021}{{\em Phys. Lett. B}
  {\bf 795} (2019)  398--423}, \href{http://arxiv.org/abs/1812.06359}{{\tt
  arXiv:1812.06359 [hep-ex]}}.

\bibitem{CMS:2018zvv}
{\bf CMS} Collaboration, A.~M. Sirunyan {\em et al.}, ``{Search for an exotic
  decay of the Higgs boson to a pair of light pseudoscalars in the final state
  with two b quarks and two $\tau$ leptons in proton-proton collisions at
  $\sqrt{s}=$ 13 TeV},''
  \href{http://dx.doi.org/10.1016/j.physletb.2018.08.057}{{\em Phys. Lett. B}
  {\bf 785} (2018)  462}, \href{http://arxiv.org/abs/1805.10191}{{\tt
  arXiv:1805.10191 [hep-ex]}}.

\bibitem{ATLAS:2018pvw}
{\bf ATLAS} Collaboration, M.~Aaboud {\em et al.}, ``{Search for the Higgs
  boson produced in association with a vector boson and decaying into two
  spin-zero particles in the $H \rightarrow aa \rightarrow 4b$ channel in $pp$
  collisions at $\sqrt{s} = 13$ TeV with the ATLAS detector},''
  \href{http://dx.doi.org/10.1007/JHEP10(2018)031}{{\em JHEP} {\bf 10} (2018)
  031}, \href{http://arxiv.org/abs/1806.07355}{{\tt arXiv:1806.07355
  [hep-ex]}}.

\bibitem{ATLAS:2020ahi}
{\bf ATLAS} Collaboration, G.~Aad {\em et al.}, ``{Search for Higgs boson
  decays into two new low-mass spin-0 particles in the 4$b$ channel with the
  ATLAS detector using $pp$ collisions at $\sqrt{s}= 13$ TeV},''
  \href{http://dx.doi.org/10.1103/PhysRevD.102.112006}{{\em Phys. Rev. D} {\bf
  102} (2020) no.~11, 112006}, \href{http://arxiv.org/abs/2005.12236}{{\tt
  arXiv:2005.12236 [hep-ex]}}.

\bibitem{CMS:2021bvh}
{\bf CMS} Collaboration, ``{Search for exotic decay of the Higgs boson into two
  light pseudoscalars with four photons in the final state at $\sqrt{s}$ = 13
  TeV},''.

\bibitem{ATLAS:2018jnf}
{\bf ATLAS} Collaboration, M.~Aaboud {\em et al.}, ``{Search for Higgs boson
  decays into pairs of light (pseudo)scalar particles in the $\gamma\gamma jj$
  final state in $pp$ collisions at $\sqrt{s}=13$ TeV with the ATLAS
  detector},'' \href{http://dx.doi.org/10.1016/j.physletb.2018.06.011}{{\em
  Phys. Lett. B} {\bf 782} (2018)  750--767},
  \href{http://arxiv.org/abs/1803.11145}{{\tt arXiv:1803.11145 [hep-ex]}}.

\bibitem{Shelton:2021xwo}
J.~Shelton and D.~Xu, ``{Exotic Higgs Decays to Four Taus at Future
  Electron-Positron Colliders},''
\newblock 10, 2021.
\newblock \href{http://arxiv.org/abs/2110.13225}{{\tt arXiv:2110.13225
  [hep-ph]}}.

\bibitem{Kato:2021ILC4b}
Y.~Kato, ``{Search for Higgs decaying to exotic scalers at the ILC},''
\newblock 11, 2021.
\newblock
  \href{http://arxiv.org/abs/https://indico.ihep.ac.cn/event/14938/}{{\tt
  https://indico.ihep.ac.cn/event/14938/}}.

\bibitem{Jung:2021tym}
S.~Jung, Z.~Liu, L.-T. Wang, and K.-P. Xie, ``{Probing Higgs boson exotic
  decays at the LHC with machine learning},''
  \href{http://dx.doi.org/10.1103/PhysRevD.105.035008}{{\em Phys. Rev. D} {\bf
  105} (2022) no.~3, 035008}, \href{http://arxiv.org/abs/2109.03294}{{\tt
  arXiv:2109.03294 [hep-ph]}}.

\bibitem{deBlas:2019rxi}
J.~de~Blas {\em et al.}, ``{Higgs Boson Studies at Future Particle
  Colliders},'' \href{http://dx.doi.org/10.1007/JHEP01(2020)139}{{\em JHEP}
  {\bf 01} (2020)  139}, \href{http://arxiv.org/abs/1905.03764}{{\tt
  arXiv:1905.03764 [hep-ph]}}.

\bibitem{ATLAS-CONF-2021-053}
{\bf ATLAS Collaboration} Collaboration, ``{Combined measurements of Higgs
  boson production and decay using up to $139$ fb$^{-1}$ of proton-proton
  collision data at $\sqrt{s}= 13$ TeV collected with the ATLAS experiment},''
  tech. rep., CERN, Geneva, Nov, 2021.
\newblock \url{https://cds.cern.ch/record/2789544}.
\newblock All figures including auxiliary figures are available at
  https://atlas.web.cern.ch/Atlas/GROUPS/PHYSICS/CONFNOTES/ATLAS-CONF-2021-053.

\bibitem{Feng:2022inv}
J.~L. Feng {\em et al.}, ``{The Forward Physics Facility at the High-Luminosity
  LHC},'' in {\em {2022 Snowmass Summer Study}}.
\newblock 3, 2022.
\newblock \href{http://arxiv.org/abs/2203.05090}{{\tt arXiv:2203.05090
  [hep-ex]}}.

\end{thebibliography}\endgroup
\end{document}